\documentclass[useAMS,usenatbib]{mn2e}
\usepackage{amssymb,graphicx,amsmath}

\newcommand{\qm}[1]
{``#1''}

\title[Quasar bolometric corrections]{Quasar bolometric corrections: theoretical considerations} 
\author[Nemmen \& Brotherton]{Rodrigo S. Nemmen,$^{1}$\thanks{E-mail: rodrigo.nemmen@ufrgs.br} Michael S. Brotherton,$^{2}$ \\
$^{1}$Instituto de F\'isica, Universidade Federal do Rio Grande do Sul, Campus do Vale, Porto Alegre, RS, Brazil \\
$^{2}$Department of Physics and Astronomy, University of Wyoming, Laramie, WY 82071}

\begin{document}

\date{Accepted 2010 June 18. Received 2010 June 17; in original form 2010 March 31}

\pagerange{\pageref{firstpage}--\pageref{lastpage}} \pubyear{2010}

\maketitle

\label{firstpage}

\begin{abstract}

Bolometric corrections based on the optical-to-ultraviolet continuum spectrum of quasars are widely used to quantify their radiative output, although such estimates are affected by a myriad of uncertainties, such as the generally unknown line-of-sight angle to the central engine. In order to shed light on these issues, we investigate the state-of-the-art models of Hubeny et al. (2000) that describe the continuum spectrum of thin accretion discs and include relativistic effects. We explore the bolometric corrections as a function of mass accretion rates, black hole masses and viewing angles, restricted to the parameter space expected for type-1 quasars. 
We find that a nonlinear relationship $\log L_{\rm bol}=A + B \log (\lambda L_\lambda)$ with $B \leq 0.9$ is favoured by the models and becomes tighter as the wavelength decreases. We calculate from the model the bolometric corrections corresponding to the wavelengths $\lambda = 1450 \; \rm \AA$, $3000 \; \rm \AA$ and $5100 \; \rm \AA$. In particular, for $\lambda = 3000 \; \rm \AA$ we find $A=9.24 \pm 0.77$ and $B=0.81 \pm 0.02$.
We demonstrate that the often-made assumption that quasars emit isotropically may lead to severe systematic errors in the determination of $L_{\rm bol}$, when using the method of integrating the ``big blue bump'' spectrum. For a typical viewing angle of $\approx 30^\circ$ to the quasar central engine, we obtain that the value of $L_{\rm bol}$ resulting from the isotropy assumption has a systematic error of $\approx 30\%$ high compared to the value of $L_{\rm bol}$ which incorporates the anisotropic emission of the accretion disc. 
These results are of direct relevance to observational determinations of the bolometric luminosities of quasars, and may be used to improve such estimates.

%In order to shed light on these issues, we have investigated the theoretical relation between the bolometric and monochromatic luminosities based on the optical-to-ultraviolet continuum spectrum of thin accretion discs. We use state-of-the-art spectral models for the thin accretion discs which include relativistic effects, and explore the parameter space of mass accretion rates, black hole masses and viewing angles expected for type-1 quasars. 

\end{abstract}


\section{Introduction}

Quasars are important and interesting as the most luminous non-exploding
objects in the universe.  They seem likely to exert a 
powerful influence on the evolution of the galaxies that host them, 
particularly through their extreme luminosities that can ionize and 
radiatively accelerate surrounding gas (e.g., \citealt{crenshaw03, begelman04, dmt05}).  Accretion onto supermassive black 
holes powers quasars, and our ability to estimate the masses of these 
black holes has significantly advanced our understanding and revealed 
relationships among quasars and their host galaxies (e.g., \citealt{ferrarese05, vester09}).

Other fundamental quasar properties include the bolometric luminosity and the
accretion rate, often expressed as the Eddington fraction (the ratio
of the bolometric luminosity to the Eddington luminosity).
The ability to determine the bolometric luminosity is often taken for 
granted, since the quasar continuum is observed directly in most cases
and simple scaling relationships are assumed (e.g., \citealt{vester09}).  How to make the bolometric
corrections is not agreed upon, however, and the uncertainties are probably
greater than those in determining black hole mass.  Understanding how
to properly make bolometric corrections and their uncertainties is
important to future quasar investigations. 

The idea behind determining bolometric luminosity is simple: from one
or more observations determine the total energy radiated by a quasar per second
into its surroundings.  For objects like stars this is relatively easy, but
that is not the case for quasars.  
First, quasars emit anisotropically, in a complicated manner, and the 
degree of this anisotropy depends on wavelength and likely varies
significantly from object to object. Second, quasars emit at all wavebands, 
from radio through X-rays and beyond, many of which are difficult to 
observe.  Third, portions of the spectrum represent reprocessed photons;
that is, some fraction of optical or UV photons can heat surrounding dust
which in turn reradiates at infrared wavelengths, so double counting
can be an issue in theoretical calculations.  Fourth, stars and stellar-heated dust can, in certain
parts of the spectrum (e.g., far infrared, optical) contaminate quasar
SEDs and be difficult to remove.

These complications make it difficult to determine the bolometric 
luminosity of even extremely well-observed individual quasars, let 
alone that of a random quasar with only limited observations.  Still,
the bolometric luminosity is an important quantity and theoretical
considerations can help us better understand the problem of bolometric 
corrections and may suggest improvements to the methods currently employed. 
 
One standard practice to determine bolometric corrections has been to 
use observations of quasar spectral energy distributions (SEDs) to determine the
apparent bolometric luminosity, by integrating over all wavelengths, and
then fitting relationships between optical magnitudes and total bolometric
luminosity (e.g., \citealt{elvis94, richards06}).  It is generally
assumed, using this approach, that the quasar is responsible for the light
being emitted at all wavelengths and that the emission is isotropic.  
The bolometric corrections derived generally show about an order of 
magnitude of dispersion around the average value. This approach 
has generally only been used to fit simple linear relationships.

Another standard practice (e.g., \citealt{netzer07}) has been to assume that 
the optical-ultraviolet portion of the quasar SED is energetically dominant 
and powers emission at other wavelengths.  This approach avoids double 
counting, and is simpler in that only a portion of the SED, the so-called \qm{big 
blue bump} thought to be emitted by a geometrically thin accretion disc, 
is important to know. Isotropic emission is generally assumed.
This technique generates bolometric corrections not too dissimilar from
those of the above technique, and apparent uncertainties are also of about an 
order of magnitude.

We can look toward accretion disc theory in an attempt to improve on these
simple approaches.  Modern disc models, such as those by \citet{hubeny00, hubeny01}, may not perfectly explain the shape of quasar big blue bumps
(e.g., \citealt{shang05}), but can provide insight into a number of 
assumptions that go into making bolometric corrections.  Below we examine
a number of features of these state-of-the-art disc models and what they
tell us about this problem.

This paper is organized as follows. Section \ref{sec:models} describes the model for the continuum spectrum emitted by thin accretion discs in type-1 quasars that we use; Section \ref{sec:angle} analyzes the dependence of the disc emission on the viewing angle and Section \ref{sec:correct} outlines how we calculate the bolometric luminosity emitted by the disc and the bolometric corrections. Section \ref{sec:results} contains the theoretical results, with \textsection \ref{sec:wave} describing the dependence of the bolometric correction on different wavelengths in the optical-UV band, \textsection \ref{sec:mdot} the corresponding dependence on the mass accretion rate and black hole mass, and \textsection \ref{sec:iso} giving an account of the systematic error obtained in the bolometric correction when quasars are assumed to emit isotropically. Section \ref{sec:disc} discusses the broader implications of our work and its possible applications to improve bolometric corrections using quasar observations. Finally, Section \ref{sec:sum} summarizes our results.

\section{Modern thin accretion disc models} \label{sec:models}

We intend to perform a self-consistent theoretical calculation of the bolometric luminosities of quasars, based on their ultraviolet-to-optical continuum spectrum (i.e. the big blue bump) and correlate it with the emission properties at different wavelengths. In order to do this, we need detailed model spectra of accretion discs, appropriate for quasars.

The underlying assumption behind the use of the big blue bump as a proxy of the bolometric luminosity is that the UV-optical band is where most of the energy in the quasars is released, whereas emission in other bands is either a product of reprocessing of the UV light (e.g. IR) or are not energetically significant on their own (e.g. radio) (e.g., \citealt{koratkar99}).

% Basic description of Hubeny model
Over the years, many models for the continuum spectra of thin accretion discs have been formulated with varying degrees of sophistication (e.g., \citealt{nt73, sun89, laor89, laor90, sincell98, hubeny00, hubeny01}). For our purposes, we adopted the detailed model of \citet{hubeny00}, which consists of a time-steady, geometrically thin, optically thick accretion disc which includes general relativistic and non-LTE effects. Comptonization is neglected since it has little effect for the range of black hole masses that we explored \citep{hubeny01}. The Hubeny et al. model calculates self-consistently the vertical structure of the disc together with the radiation field. It was shown by \citet{shang05} to reproduce some observed properties of the UV spectra of quasars and is also supported by near-IR observations \citep{kishimoto08}. These models may not be perfect, but they are currently the best theory has to offer.

\subsection{Parameter space explored} \label{sec:pars}

\begin{figure}
\centering
\includegraphics[scale=0.5]{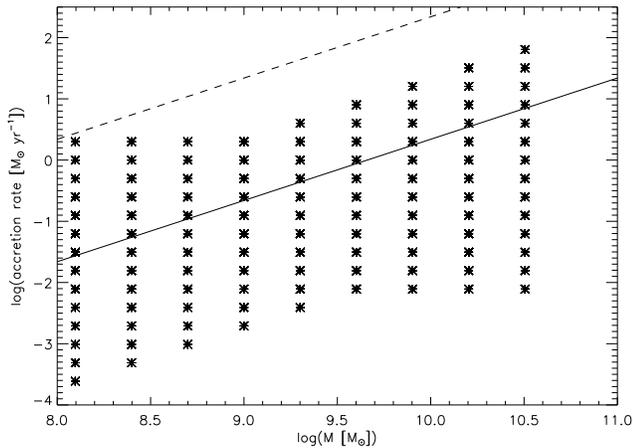}
\caption{$\dot{M}$ vs. $M$ parameters of the Hubeny et al. grid of models for the Kerr case. The solid line illustrates the relation $\dot{m} = 0.01$ and the dashed line corresponds to $\dot{m} = 1$.}
\label{fig:pars}
\end{figure}

% Parameter space explored
\citet{hubeny00} constructed a grid of models for a wide range of values of black hole mass $M$ and mass accretion rate $\dot{M}$, for two values of the viscosity parameter $\alpha$ (0.01 and 0.1) and for the two extreme values of the black hole spin $a$ corresponding to a maximally spinning Kerr black hole ($a/M=0.998$) and a nonrotating Schwarzschild hole ($a/M=0$). 
The black hole masses explored range from $1.25 \times 10^8 M_\odot$ to $3.2 \times 10^{10} M_\odot$; the accretion rates range from $2.4 \times 10^{-4} \; M_\odot \ {\rm yr}^{-1}$ to $64 \; M_\odot \ {\rm yr}^{-1}$, or expressed in terms of the Eddington accretion rate, between $1.1 \times 10^{-5} \dot{M}_{\rm Edd}$ and $0.7  \dot{M}_{\rm Edd}$ where the Eddington accretion rate is defined as 
$\dot{M}_{\rm Edd} \equiv 22 M /(10^{9} M_\odot) \; M_\odot \, {\rm yr}^{-1}$. The maximal luminosity attained by the models correspond to $L/L_{\rm Edd} \approx 0.3$. Figure \ref{fig:pars} shows the $\dot{M}$ vs. $M$ parameters of the Hubeny et al. grid of models for the Kerr case.
Hubeny et al. also explored the effect of the inclination of the disc with respect to the line of sight on the observed spectra, exploring angles ranging from nearly face-on ($\theta \approx 0^\circ$) to nearly edge-on discs ($\theta \approx 90^\circ$).

% QSO regime, restriction to type-1 AGN
In our exploration of the parameter space of the accretion disc model, we restrict ourselves to non-LTE models and to accretion rates expected to be appropriate for quasars.
It is generally accepted that depending on amount of mass supplied to the central black hole, different accretion flow states are possible. For instance, it is thought that in the regime of accretion rates appropriate for bright AGNs and quasars, $\dot{m} \gtrsim 0.01$, the flow is in the thin disc state, while for $\dot{m} \lesssim 0.01$ -- the regime appropriate for low-luminosity AGNs such as LINERs -- the flow would be in a radiatively inefficient or advection-dominated state (ADAF; e.g., \citealt{narayan08}), where we define the dimensionless accretion rate $\dot{m} \equiv \dot{M}/\dot{M}_{\rm Edd}$. In our analysis we consider only accretion rates with $\dot{m} \geq 0.01$. In Figure \ref{fig:pars}, we illustrate the relation $\dot{m} = 0.01$ as the solid line and $\dot{m} = 1$ as the dashed line.

% plot produced with pars.pro

\subsection{The impact of the inclination angle} \label{sec:angle}

Optically thick, geometrically thin accretion discs emit light anisotropically, so when calculating their total amount of power radiated away we must take that aspect into account. For a Newtonian disc, the dependency of the specific flux on the inclination of the disc with respect to the line of sight $\theta$ is simply $F_\nu \propto \cos \theta$ \citep{frank02}. The disc model of \citet{hubeny00} takes into account general relativistic effects such as beaming, aberration and light bending as well as limb darkening, which introduce complicated angle-dependencies in the observed spectrum.

\begin{figure}
\centering
\includegraphics[scale=0.7]{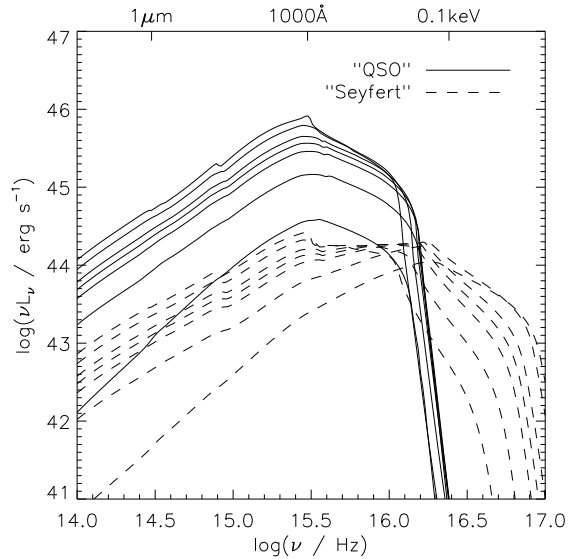}
\caption{Theoretical spectra computed with the model of \citet{hubeny00} with parameters typical of quasar accretion discs ($M=10^9 M_\odot$, $\dot{M}=0.25 \; M_\odot \ {\rm yr}^{-1}$; \textit{solid line}) and bright Seyferts ($M=1.25 \times 10^8 M_\odot$, $\dot{M}=0.03 \; M_\odot \ {\rm yr}^{-1}$; \textit{dashed line}). Each set of models differ only with respect to the viewing angle, where the lowermost spectrum in each case corresponds to the nearly edge-on angle and the uppermost spectrum represents the nearly face-on orientation (see text for the values of $\theta$).}
\label{fig:seds}
\end{figure}

To illustrate this point, Figure \ref{fig:seds} shows the spectra of models which differ only with respect to the viewing angle, with all other parameters held fixed. The specific luminosity is calculated as the one an observer along a particular viewing angle would see if the source were isotropic, i.e. we use the equation $L_\nu = 4 \pi d^2 F_\nu$ where $d$ is the distance. Two sets of models are displayed in that figure, with parameters appropriate for quasars -- $M=10^9 M_\odot$, $\dot{M}=0.25 \; M_\odot \ {\rm yr}^{-1}$ -- and bright Seyferts --  $M=1.25 \times 10^8 M_\odot$, $\dot{M}=0.03 \; M_\odot \ {\rm yr}^{-1}$ -- both sets with $\alpha=0.1$ and $\dot{m} \approx 0.01$. The viewing angles in each set of spectra range from the edge-on (lowermost spectrum in each case) to the face-on case (uppermost spectrum), with models computed for the following values of $\theta$: $89.4^\circ$, $78.5^\circ$, $66.4^\circ$, $60^\circ$, $53^\circ$, $36.9^\circ$ and $8.1^\circ$. Figure \ref{fig:seds} illustrates that the viewing angle has a pronounced impact on the spectra of the thin disc.

A more insightful way of exhibiting the viewing angle dependency of the models can be obtained by plotting the integrated luminosity emitted by the disc as a function of viewing angle. We compute the integrated luminosity as
\begin{equation} \label{eq:li}
L_i = \int_{\nu_0}^{\nu_1} L_\nu \; d\nu,
\end{equation}
where $\nu_0$ and $\nu_1$ are respectively the lowest and highest frequencies at which the thin disc radiates. $L_i$ will obviously have the same $\theta$-dependency as $F_\nu$.

\begin{figure}
\centering
\includegraphics[scale=0.7]{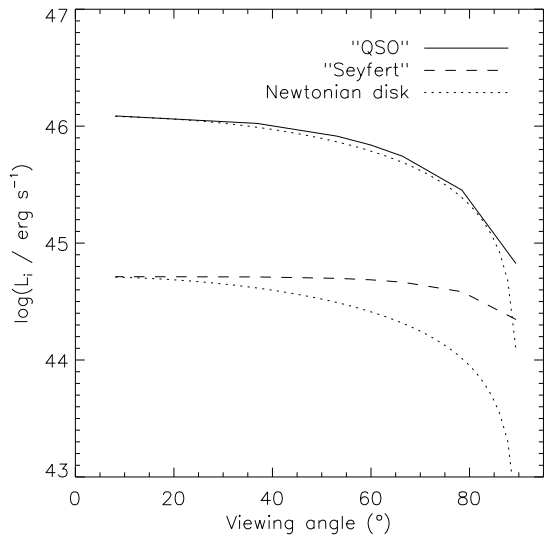}
\caption{Dependency of the integrated luminosity $L_i$ of the thin accretion disc on the viewing angle $\theta$. The solid line corresponds to the ``quasar'' fiducial parameters used in Figure \ref{fig:seds} and the dashed line represents the ``Seyfert'' fiducial parameters. The dotted line illustrates the $\theta$-dependency of the simple Newtonian thin disc model for which $F_\nu \propto \cos \theta$. Note the log scale.} 
\label{fig:angle}
\end{figure}

Figure \ref{fig:angle} shows how $L_i$ depends on $\theta$ for the two fiducial sets of model spectra mentioned previously (and plotted in Figure \ref{fig:seds}), compared to the Newtonian case in which $F_\nu \propto \cos \theta$. Clearly, the relativistic model of \citet{hubeny00} show significant departures from the simple Newtonian accretion model, especially at large inclination angles. Such anisotropic emission must be taken into account when estimating the total luminosity of the thin disc and therefore the bolometric luminosity of quasars.

% plot produced with angle.pro

\subsection{Theoretical calculation of the bolometric correction} \label{sec:correct}

The total luminosity radiated by the disc is independent of the viewing angle and is calculated from $L_i$ as
\begin{eqnarray} \label{eq:l}
L & = &  \frac{1}{4\pi} \int_{0}^{2\pi} \int_{0}^{\pi}  L_i(\theta) \sin \theta \; d\theta \; d\phi \nonumber \\
& = & \int_{0}^{\pi/2}  L_i(\theta) \sin \theta \; d\theta,
\end{eqnarray}
since the grid of models produced by \citet{hubeny00} was calculated ranging up to the angle $\theta \approx 90^\circ$.

As mentioned before, a standard practice to estimate the bolometric luminosity from quasar SEDs is to use the optical-UV portion of the spectrum on the 
assumption that it is the energetically dominant component of the radiative output and powers the emission at other wavelengths (e.g., \citealt{netzer07}). We follow this principle here to estimate $L_{\rm bol}$ and simply assume 
\begin{equation} \label{eq:lbol} 
L_{\rm bol} = L, 
\end{equation}
where $L$ is calculated using Equation \ref{eq:l}.

\begin{figure}
\centering
\includegraphics[scale=0.7]{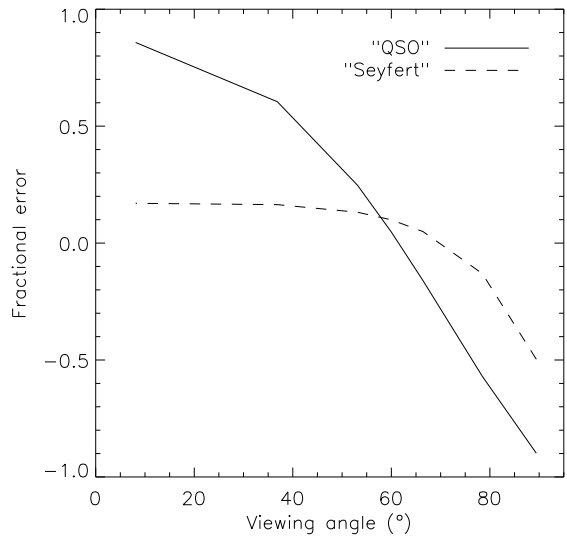}
\caption{Fractional error $(L_i-L)/L$ incurred when the bolometric luminosity is estimated assuming that the accretion disc is an isotropic radiator, as a function of viewing angle, for the quasar and the Seyfert fiducial models (parameters listed in \textsection \ref{sec:angle}).}
\label{fig:angleerror}
\end{figure}

This formulation for computing $L_{\rm bol}$ has the virtue that it accounts for the anisotropic nature of the emission of the thin accretion disc, including all the relevant effects. When computing the luminosity of the thin disc, many
studies have assumed that the disc emits isotropically which introduces errors. Figure \ref{fig:angleerror} shows the fractional error incurred when the bolometric luminosity is estimated assuming isotropic emission, which corresponds to using the equation $L_{\rm bol} = L_i$ as opposed to equation \ref{eq:lbol}, for both the quasar and the Seyfert fiducial models mentioned before. The fractional error plotted corresponds to $(L_i-L)/L$. We can see in this figure that errors in the estimate of $L_{\rm bol}$ as high as $\approx 90\%$ for the face-on case can occur when the thin disc is wrongly assumed to radiate isotropically. 

% plot produced with angleerror.pro

\section{Results} \label{sec:results}

In this section, we explore the different results we are able to obtain from the grid of models for quasar spectra described in \textsection \ref{sec:models}. We are concerned particularly about the resulting bolometric luminosities and their dependence on the properties of the central engine such as black hole mass, mass accretion rate and viewing angle. We will also explore the relation of $L_{\rm bol}$ to the monochromatic luminosity at different wavelengths in the optical-UV band.

\subsection{Dependence of bolometric correction on wavelength} \label{sec:wave}
		
% Histograms
One quantity that is of particular interest is the ``bolometric correction'' $\zeta_\lambda \equiv L_{\rm bol}/(\lambda L_\lambda)$, which provides a way of estimating the bolometric luminosity knowing the monochromatic luminosity at a certain wavelength $\lambda$ in the optical-UV band. 
According to the unification scenario of AGNs, the optical-UV emission would only be directly observable in type-1 objects (e.g., \citealt{bar89}, \citealt{anton93}). Therefore we restrict our study of the QSO synthetic spectra to this class of objects. We do so by restricting the viewing angles that we consider to $\theta \leq 60^\circ$, which we call ``type-1'' models.  We note that the
opening angle may differ from this value, and that it itself may be 
dependent on luminosity (e.g., receding torus models, e.g., \citealt{law91};
\citealt{sim05}) or other parameters. 
 
It is worthwhile to analyse the distribution of values of $\zeta_\lambda$ for different regions of the parameter space, for different values of $\lambda$. We calculate the values of $\zeta_\lambda$ for three wavelengths: $1450 \; \rm \AA$, $3000 \; \rm \AA$ and $5100 \; \rm \AA$. Flux measurements at these wavelengths are often adopted in observational studies to make estimates of bolometric luminosities \citep{elvis94,kaspi00, richards06}. The wavelengths $1450 \; \rm \AA$, $3000 \; \rm \AA$ and $5100 \; \rm \AA$ correspond to spectral regions in the optical for high-redshift, intermediate-redshift and low-redshift quasars respectively, and at these wavelengths the emission lines are not present or are relatively weak. 

\begin{figure*}
\centering
\includegraphics[scale=1]{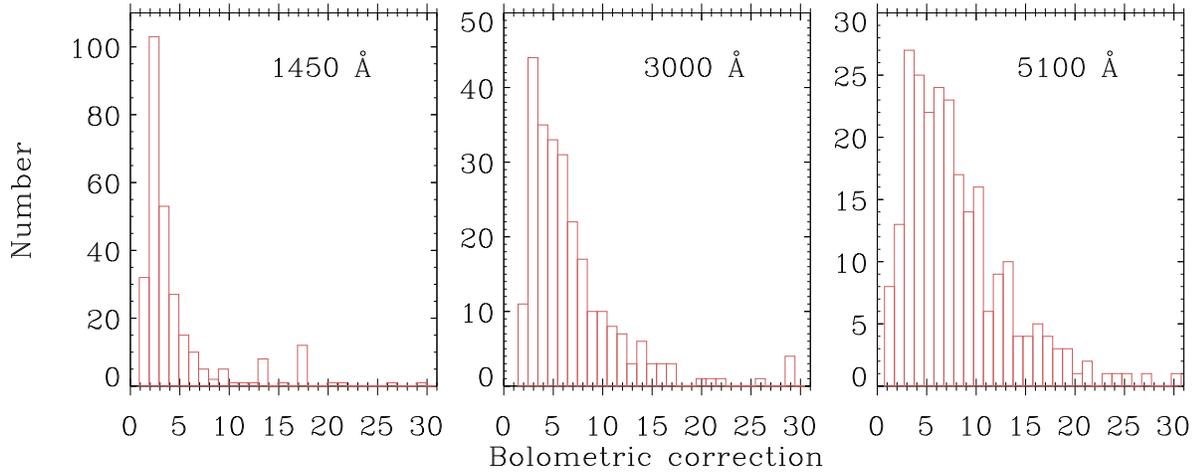}
\caption{Histograms of the distribution of values of the bolometric correction $\zeta_{\lambda}$ for the wavelengths $1450 \; \rm \AA$, $3000 \; \rm \AA$ and $5100 \; \rm \AA$, for the ``type-1'' quasar models.}
\label{fig:hist}
\end{figure*}

Figure \ref{fig:hist} shows the histograms of the distribution of values of $\zeta_\lambda$ for non-LTE models corresponding to ``type-1'' models with $\dot{m} \geq 0.01$, i.e. ``type-1 quasars''. We can see in Figure \ref{fig:hist} that as the wavelength increases, the distribution of values of $\zeta_\lambda$ broadens. The peak of the distribution moves towards higher values as $\lambda$ increases, with the medians of the bolometric correction for each wavelength being $\left \langle \zeta_{1450} \right \rangle = 3$, $\left \langle \zeta_{3000} \right \rangle = 5.9$ and $\left \langle \zeta_{5100} \right \rangle = 7.6$.

% plot produced with hist.pro

\begin{figure}
\centering
\includegraphics[scale=0.5]{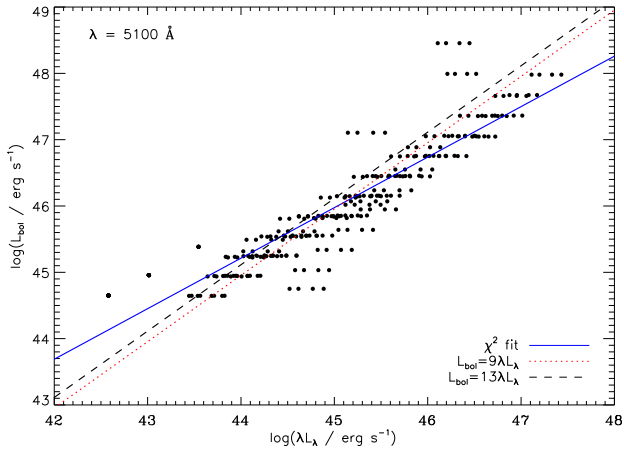}
\caption{Relation between $L_{\rm bol}$ and $\lambda L_\lambda$ at $\lambda=5100 \; \rm \AA$, for the ``type-1'' quasar model spectra. The dotted and dashed lines correspond to models of the form $L_{\rm bol} = \zeta_{5100} \lambda L_\lambda$, while the solid line show the best-fitted power-law to the data using the $\chi^2$ method.}
\label{fig:5100}
\end{figure}

% 5100 A
Figure \ref{fig:5100} shows the relation between $L_{\rm bol}$ and $\lambda L_\lambda$ with $\lambda=5100 \; \rm \AA$, for the type-1 quasar models with $\dot{m} \geq 0.01$. The dotted and dashed lines in Figure \ref{fig:5100} correspond to models in which the bolometric luminosity is simply related to the monochromatic luminosity at $5100 \; \rm \AA$ through $L_{\rm bol} = \zeta_{5100} \lambda L_\lambda$, with the value of $\zeta_{5100}$ being respectively 9 and 13 for the dotted and dashed lines. These values of $\zeta_{5100}$ are the ones derived by \citet{kaspi00} and \citet{elvis94} respectively. The mean deviation $\sigma (\log L_{\rm bol})$ of the data about these lines is 0.47 ($\zeta_{5100} = 9 \times$ model) and 0.49 ($\zeta_{5100} = 13 \times$ model). 

We parametrize a power-law fit in the log-log space portrayed in Figure \ref{fig:5100} as $\log L_{\rm bol}=A+B \log (\lambda L_\lambda)$ and fit it to the data using the $\chi^2$ statistic, obtaining the parameters $A=11.7 \pm 0.93$ and $B=0.76 \pm 0.02$ (solid line), with a mean deviation about the best-fit model of 0.38. This result suggests that the non-linear model  of the form $L_{\rm bol} \propto(\lambda L_\lambda)^{0.76}$ is favoured to explain the data for the sample of simulated type-1 quasar spectra for $\lambda = 5100 \; \rm \AA$.

% plot produced with nulumbol.pro

\begin{figure}
\centering
\includegraphics[scale=0.5]{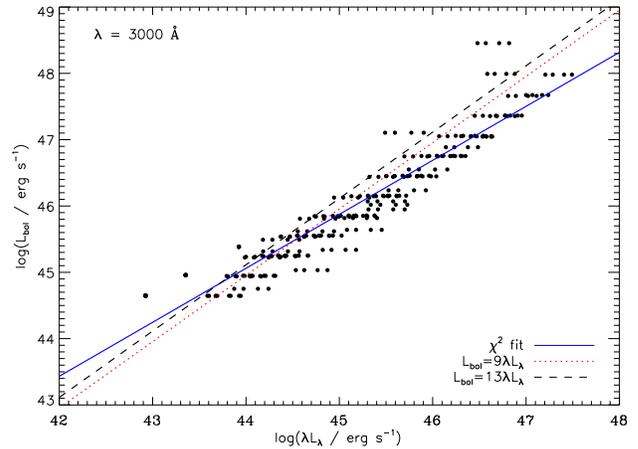}
\caption{Same as Figure \ref{fig:5100} for $\lambda=3000 \; \rm \AA$.}
\label{fig:3000}
\end{figure}

% 3000 A
Figure \ref{fig:3000} shows the $L_{\rm bol}$ vs. $\lambda L_\lambda$ for type-1 quasar models at $\lambda=3000 \; \rm \AA$, following the same notation as Figure \ref{fig:5100}. A fit to the $3000 \; \rm \AA$ data using the $\chi^2$ method results in $A=9.24 \pm 0.77$ and $B=0.81 \pm 0.02$ (dot-dashed line), with a mean deviation about the best-fit model of 0.31. For comparison, the corresponding mean deviations about the linear models with $\zeta_{3000}=9$ and $\zeta_{3000}=13$ are respectively 0.38 and 0.46. 

% plot produced with nulumbol.pro

\begin{figure}
\centering
\includegraphics[scale=0.5]{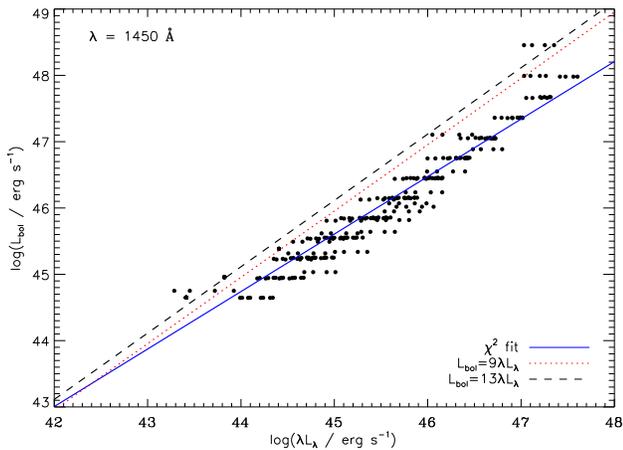}
\caption{Same as Figure \ref{fig:5100} for $\lambda=1450 \; \rm \AA$.}
\label{fig:1450}
\end{figure}

% 1450 A
Figure \ref{fig:1450} shows the $L_{\rm bol}$ vs. $\lambda L_\lambda$ for type-1 quasar models at $\lambda=1450 \; \rm \AA$, again following the notation of Figure \ref{fig:5100}. A fit to the $1450 \; \rm \AA$ data using the $\chi^2$ method results in $A=6.7 \pm 0.69$ and $B=0.87 \pm 0.02$, with a mean deviation about the best-fit model of 0.26. For comparison, the mean deviations about the linear models with $\zeta_{1450}=9$ and $\zeta_{1450}=13$ are respectively 0.51 and 0.64. 

% plot produced with nulumbol.pro

\subsection{Dependence of bolometric correction on the accretion rate and black hole mass} \label{sec:mdot}

In this section we analyze the possible dependence of the bolometric correction on the Eddington ratio $L_{\rm bol}/L_{\rm Edd}$, which traces the accretion rate, and the black hole mass for the ``type-1'' quasar model spectra. 

\begin{figure}
\centering
\includegraphics[scale=0.5]{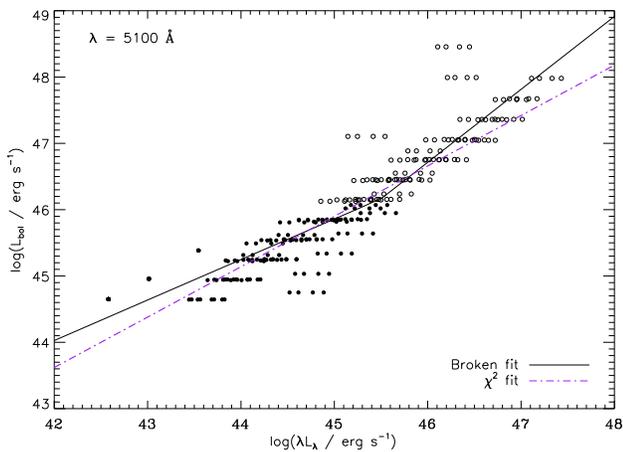}
\caption{Relation between $L_{\rm bol}$ and $\lambda L_\lambda$ at $\lambda=5100 \; \rm \AA$, for the ``type-1'' quasar model spectra. The filled circles correspond to $L_{\rm bol}/L_{\rm Edd} < 0.1$ and the open circles correspond to $L_{\rm bol}/L_{\rm Edd} \geq 0.1$. The solid and dot-dashed lines correspond to a broken power-law fit and a simple power-law, respectively.}
\label{fig:eddrat}
\end{figure}

Figure \ref{fig:eddrat} shows the relation between $L_{\rm bol}$ and $\lambda L_\lambda$ at $\lambda = 5100 \; \rm \AA$, where the filled circles correspond to $L_{\rm bol}/L_{\rm Edd} < 0.1$ and the open circles correspond to $L_{\rm bol}/L_{\rm Edd} \geq 0.1$. There is a hint in this plot that the slope of the dependence of $L_{\rm bol}$ on $\lambda L_\lambda$ changes slightly as the Eddington ratio increases, with a break at $\log (\lambda L_\lambda) \sim 45.5$ corresponding to the chosen ``transition value'' $L_{\rm bol}/L_{\rm Edd} = 0.1$. The solid line in Figure \ref{fig:eddrat} displays the broken power-law best-fitted to the data using the Marquardt-Levenberg least-squares method, with the break at $x=45.5$. The mean deviation of the data about this best-fit model is 0.35.

% plot produced with nonlinear.pro

\begin{figure}
\centering
\includegraphics[scale=0.5]{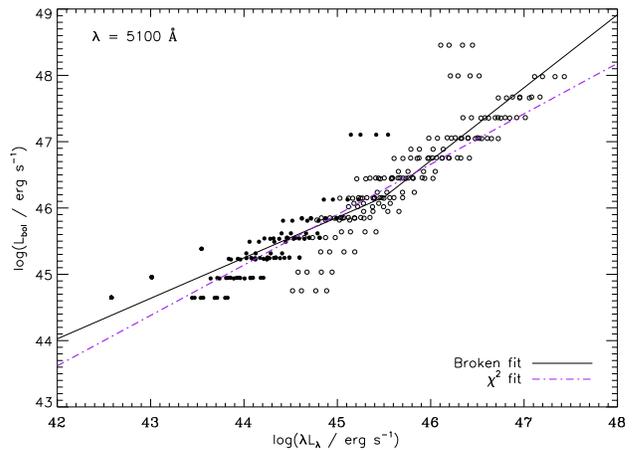}
\caption{Same as Figure \ref{fig:eddrat}, but here the filled circles correspond to $M < 10^9 M_\odot$ and the open circles correspond to $M \geq 10^9 M_\odot$.}
\label{fig:mass}
\end{figure}

Figure \ref{fig:mass} is similar to the previous one, but now the filled circles correspond to models for which $M < 10^9 M_\odot$ while the open circles represent models with $M \geq 10^9 M_\odot$. We find a slight hint in this figure of a change in the slope of the power-law fit as the black hole mass increases, similarly to what was observed in the previous plot. The solid line shows the same broken power-law fit shown in Figure \ref{fig:eddrat}.

% plot produced with nonlinear.pro

There is no strong reason to abandon a simple power-law model in favour of more complicated models such as a broken power-law, since the latter reproduces only marginally better the results compared to a simple power-law. 
 
\subsection{Beyond the isotropy assumption when computing quasar luminosities} \label{sec:iso}

When converting quasar nuclear line-of-sight fluxes to luminosities or when computing bolometric corrections and luminosities, the usual assumption is that quasars emit isotropically (e.g., \citealt{elvis94, shang05, richards06}). We know that this is not the case, as we showed in \textsection \ref{sec:correct} (c.f. Figure \ref{fig:angleerror}). Using the spectral models for quasar accretion discs, we can have a theoretical perspective on how good (or bad) is the isotropy assumption on average.

\begin{figure}
\centering
\includegraphics[scale=0.7]{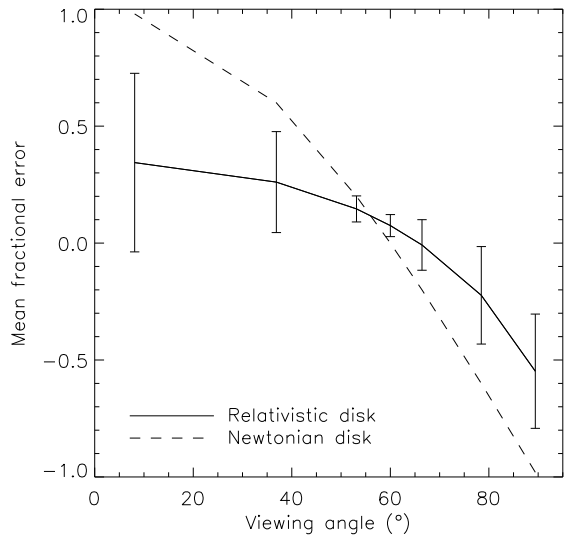}
\caption{Average fractional error incurred when the bolometric luminosity is estimated assuming that the accretion disc is an isotropic radiator, as a function of viewing angle.}
\label{fig:meanerror}
\end{figure}

In order to address this issue, we calculated the average error implied by the isotropy assumption when estimating the bolometric luminosities of accretion discs from line-of-sight fluxes measured in the optical-UV continuum spectrum. We did this by averaging all the type-1 quasar models at specific viewing angles. Figure \ref{fig:meanerror} shows the corresponding mean fractional error calculated in this way, defined as 
\begin{equation}
\textrm{Mean fractional error} = \left \langle \frac{L_i - L_{\rm bol}}{L_{\rm bol}} \right \rangle_\theta,
\end{equation}
(see equations \ref{eq:li} and \ref{eq:lbol}) as a function of the viewing angle $\theta$. We show in this figure the fractional error that result when two different accretion disc models are used: the general relativistic model (together with the $1\sigma$ uncertainty for the relativistic model) of \citet{hubeny00} and for comparison a simple Newtonian disc \citep{frank02}. 
For the Newtonian model, the bolometric luminosity is related to the integrated luminosity assuming isotropy $L_i$ as $L_{\rm bol} = 1/(2 \cos \theta) L_i$, and the fractional error is given simply by $2 \cos \theta -1$. For the general relativistic model, we calculate the mean error numerically.

For each accretion disc model there is a specific line-of-sight $\theta_i$ at which $L_{\rm bol} = L_i$, i.e. at these viewing angles the mean error is zero and there is no error incurred in using the isotropy assumption. For the Newtonian case, this angle is simply $\theta_i = 60^\circ$. For the relativistic case, Figure \ref{fig:meanerror} tells us that $\theta_i \approx 66^\circ$, with the uncertainty in the mean error being lowest at $\theta \approx 55^\circ$.

% plot produced with meanerror.pro

\section{Discussion} \label{sec:disc}

% Scatter in the models similar to scatter in real data
We note that the three main underlying assumptions behind our work are (1) that the models for the spectra of thin accretion disc correctly explain the observed optical-UV continuum spectra of quasars, (2) the parameter space for the accretion disc models that we explored closely resemble that of real quasars, and (3) the X-ray contribution to the bolometric luminosity is small compared to that of the big blue bump.
The resulting scatter of $\sigma(\log L_{\rm bol}) \approx 0.9$ dex in the model data (c.f. Figure \ref{fig:5100}) arises from the variation in the different parameters of the central engine: accretion rate, black hole mass and spin, viewing angles and viscosity. The fact that the theoretical scatter is similar to the scatter measured in observed quasar data \citep{elvis94, shang05, richards06}  
suggests that our assumptions 1 and 2 are reasonable and offers an explanation to the observed scatter.

There is accumulating evidence that less than $\approx 3 \%$ of the radiative output of quasars is released in X-rays (\citealt{elvis94,vasu07,ho08,vasu09}). This indicates that our third assumption is quite plausible.

% Non-linear models agree better with the synthetic data
Previous results based on observed quasar SEDs find simple linear bolometric corrections of the form $L_{\rm bol} \propto \lambda L_\lambda$ for different reference wavelengths \citep{elvis94, kaspi00, richards06}. Contrary to that, we find that non-linear bolometric corrections of the form $\log L_{\rm bol}=A+B \log (\lambda L_\lambda)$ better reproduces the synthetic data resulting from the accretion disc models, with $B \leq 0.9$ for the wavelengths in the optical-UV band that we considered. 
Our findings are consistent with the observations which show that the quasar SED shape changes with luminosity \citep{steffen06}.
On the other hand, we found only weak evidence for evolution of the bolometric correction with the Eddington ratio and the black hole mass.

% Smaller wavelengths -> tighter bolometric corrections
We carried our calculations of the bolometric corrections for three fiducial wavelengths, $1450 \; \rm \AA$, $3000 \; \rm \AA$ and $5100 \; \rm \AA$. We find that as the wavelength decreases, the distribution of values of $L_{\rm bol}/(\lambda L_\lambda)$ becomes narrower (c.f. Figure \ref{fig:hist}). Furthermore, the relation between $L_{\rm bol}$ and the monochromatic luminosities becomes tighter, in the sense that both the uncertainties in the parameters of the power-law fits and the dispersion around the best-fit model decrease (\textsection \ref{sec:wave}). This result simply reflects the fact that monochromatic luminosities measured closer to the peak of the energy distribution tend to be better tracers of the total energy released by the accretion disc.
% Dust matters
Although it is preferable to use luminosities emitted at increasingly smaller wavelengths for bolometric corrections, there is one important practical caveat as we advance upwards in energy: the effects of dust extinction become progressively more strong, which obviously must be considered. 

% Now, recommendations
We can make some recommendations based on the models for the spectra of accretion discs that we studied, which can be used to estimate bolometric luminosities from monochromatic luminosities at specific wavelengths. 
For $\lambda = 1450 \; \rm \AA$, the following bolometric correction formula is suggested from the models (see \textsection \ref{sec:wave}):
\begin{equation}
\log L_{\rm bol}=(6.7 \pm 0.69) + (0.87 \pm 0.02) \log L_{1450}.
\end{equation}
For $\lambda = 3000 \; \rm \AA$, we derived the correction
\begin{equation}
\log L_{\rm bol}=(9.24 \pm 0.77) + (0.81 \pm 0.02) \log L_{3000}.
\end{equation}
Finally, for $\lambda = 5100 \; \rm \AA$, the following correction is indicated:
\begin{equation}
\log L_{\rm bol}=(11.7 \pm 0.93) + (0.76 \pm 0.02) \log L_{5100}.
\end{equation}
Since the bolometric corrections are tighter for $\lambda = 1450 \; \rm \AA$, it is preferable to use the measurements at this wavelengths. When using these bolometric corrections, the readers should be aware of the underlying assumptions of our work described previously.

% Correcting isotropic estimates
We demonstrated that the widespread assumption that quasars emit isotropically leads to considerable errors in the determination of the actual bolometric luminosities, with average fractional errors exceeding 30\%.
How can these theoretical results be conceivably applied in a real-world case, to improve observational estimates of the quasar bolometric luminosity? Let us assume that the bolometric luminosity $L_{\rm bol}^{\rm iso}$ is available from integrating an observed quasar optical-UV spectrum adopting isotropy, but the line-of-sight to the quasar is unknown (as is usually the case). 
If all lines of sight to the central engine are in principle available (i.e. no dusty torus), then the average viewing angle for a quasar is $\langle \theta \rangle \approx 57^\circ$. Figure \ref{fig:meanerror} tells us that the mean \textit{systematic} fractional error is 0.1 and the actual bolometric luminosity would be systematically smaller than the isotropic estimate by a factor of $\approx 0.9$.

Of course, we must consider the presence of a dusty torus blocking our view to a large part of the accretion disc. If we consider a dusty structure such that only viewing angles $\theta \lesssim 60^\circ$ are accessible, then $\langle \theta \rangle \approx 39^\circ$. In this case, the systematic average fractional error resulting from the isotropic estimate is 0.24, such that $L_{\rm bol} \approx 0.8 L_{\rm bol}^{\rm iso}$. 
A somewhat larger dusty torus that restricts the lines of sight to $\theta \lesssim 45^\circ$ would correspond to $\langle \theta \rangle \approx 30^\circ$; the corresponding average fractional error would be 0.28 with $L_{\rm bol} \approx 0.78 L_{\rm bol}^{\rm iso}$. It is important to keep in mind these systematic errors when estimating quasar bolometric luminosities assuming isotropy.

In the case of receding torus models \citep{law91, sim05}, which have larger opening angles with
increasing quasar luminosity, edge-on discs become more commonly seen
in the most luminous systems.  In this case, bolometric corrections would be
expected to be a function of luminosity.  Additionally, the isotropy
assumption would be less in error for a larger fraction of luminous quasars,
on average.

Our results indicate that determining torus opening angles and their
distribution as a function of luminosity would help improve bolometric 
corrections.  Also finding ways of determining disc inclination angles,
perhaps through radio properties \citep{wills95} or other approaches 
\citep{down10} is also indicated as a way of significantly improving bolometric 
corrections in individual objects.

\section{Summary}	\label{sec:sum}

We have studied state-of-the-art models for the continuum spectrum of thin accretion discs, exploring the theoretical optical-to-UV bolometric corrections for 
a range of mass accretion rates, black hole masses and viewing angles appropriate for the case of type-1 quasars. We find a number of results of direct relevance in observational studies aimed at the determination of quasar bolometric luminosities.

Regarding the relation between the bolometric and the monochromatic luminosities measured in a given wavelength in the optical-to-UV band, we obtain:
\begin{enumerate}
\item A nonlinear relationship of the form $L_{\rm bol} \propto (\lambda L_{\lambda})^{\rm const}$ is favoured by the models, with the deviation from linearity increasing as the wavelength decreases.
\item The bolometric corrections become tighter with decreasing wavelength, i.e. as we approach the peak of the quasar SED (although reddening affects
shorter wavelengths more strongly).
\item We list fitting formulas for the bolometric correction corresponding to the wavelengths $\lambda = 1450 \; \rm \AA$, $3000 \; \rm \AA$, and $5100 \; \rm \AA$, which can be used to improve observational estimates.
\end{enumerate}

Quasars are not isotropic emitters, and making this assumption may lead to severe systematic errors in the determination of $L_{\rm bol}$ using the method of integrating the ``big blue bump'' spectrum.
For a typical line-of-sight of $\approx 30^\circ$ or less to the quasar central engine, we estimate that there is a systematic error of $\approx 30\%$ or higher implied if $L_{\rm bol}$ is estimated using the isotropy assumption. We find that a more careful calculation should be $L_{\rm bol} \approx 0.8 L_{\rm bol}^{\rm iso}$, where $L_{\rm bol}^{\rm iso}$ is the bolometric luminosity estimated from integrating the optical-to-UV spectrum assuming isotropy.

With the results of this work, a better assessment of the radiative output of quasars based on observations will be made possible.

\section*{Acknowledgments}

We thank Omer Blaes for making the library of thin disk spectral models available to the community. RSN is grateful to the hospitality of Department of Physics \& Astronomy at the University of Wyoming, were part of this work was carried out, and acknowledges the financial support of CNPq. MB is grateful to the hospitality of Instituto de F\'isica, Universidade Federal do Rio Grande do Sul in Porto Alegre, Brazil, and acknowledges support from NASA through grant NNG05GE84G.

%\appendix

\bsp

\label{lastpage}

\end{document}